# Hypersensitive Transport in Photonic Crystals with Accidental Spatial Degeneracies

Eleana Makri[1], Kyle Smith[2], Andrey Chabanov[2], Ilya Vitebskiy[3], Tsampikos Kottos[1]

[1]*Department of Physics,Wesleyan University, Middletown, CT-06459, USA*
[2]*Departmet of Physics and Astronomy, University of Texas at San Antonio, TX-78249, USA*
[3]*Air Force Research Laboratory, Sensors Directorate, Wright Patterson Air Force Base, OH-45433, USA*

A localized defect mode in a photonic-layered structure develops nodal points. Placing a thin metallic layer at such a nodal point results in the phenomenon of induced transparency. We demonstrate that if this nodal point is not a point of symmetry, then even a tiny alteration of the permittivity in the vicinity of the defect suppresses the localized mode along with the resonant transmission; the layered structure becomes highly reflective within a broad frequency range. Applications of this hypersensitive transport for optical limiting and switching are discussed.

One of the main technological and fundamental challenges of our days is the design of photonic structures that allow for an efficient manipulation of the amplitude, phase, polarization, or direction of electromagnetic signals [1]. Successful management of these features can lead to many diverse applications ranging from microwave and optical communications, sensors and power limiters, to energy harvesting, switching, and optical computing [2,3]. In this endeavor, the control of the interaction between electromagnetic radiation and matter is of utmost importance.

An efficient way to achieve this control is via localized modes supported by the photonic structure. These modes develop nodal points where the electric field is infinitesimally small. Placing a thin metallic nanolayer at these positions will not affect the localized mode and the resonance transmission associated with this mode. This is the well-known phenomenon of induced transparency [4]. When a small alteration $\Delta\varepsilon$ in the permittivity of a nearby layer(s) occurs it will affect the localized mode. Depending on the nodal point symmetry, there are three possible scenarios: (a) the nodal point coincides with the mirror plane of the entire layered structure before and after the perturbation; (b) the nodal point coincides with the mirror plane in the original configuration, but the perturbation breaks this symmetry and (c) the nodal point of the localized mode is not a symmetry point, in which case the coincidence of the metallic nanolayer and the node of localized mode can be viewed as *accidental spatial degeneracy* (ASD). In case (a), the symmetric alteration of the layered structure results in a simple shift of the localized mode frequency. The metallic nanolayer still coincides with the nodal point of the modified localized mode and hence does not affect the resonant transmission. In the cases (b) and (c), the nodal point of the localized mode shifts away from the metallic nanolayer, which can result in a dramatic suppression of the localized mode, along with the resonant transmission. In either case, the entire layered structure becomes opaque at any frequency. Due to the presence of metallic nanolayer, the abrupt transition from resonant transparency to broadband opacity can be caused by just a tiny change in the defect's permittivity, which justifies the use of the term *hypersensitivity*. The above feature equally applies to the cases (b) and (c), but with one important exception, when the permittivity alteration is self-induced by the localized mode. Typically, a self-induced change in the permittivity is associated with nonlinear effects, heating, etc. If the permittivity alteration $\Delta\varepsilon$ is indeed self-induced; the transition from resonant transparency to broadband opacity is much more pronounced and abrupt in the case (c) of accidental spatial degeneracy, as compared to the case (b), where the undisturbed layered structure was symmetric.

In most of the existing applications of metallo-dielectric layered structures (see for example [4,5]) the abovementioned hypersensitivity of asymmetric configurations to a self-induced alteration of the refractive index would be undesirable and counterproductive. In this paper we take an alternative viewpoint. We demonstrate how such hypersensitivity can be used in microwave and optical limiters, and show that it can dramatically enhance the device performance. As an example, we consider a microwave limiter based on asymmetric metal-dielectric layered structure supporting localized mode with ASD. We show that even a small self-induced alteration of the refractive index at the location of the localized mode produces an abrupt transition from resonant transparency for low-level radiation to high broadband reflectivity for high-level radiation. If, on the other hand, the asymmetric permittivity alteration is not self-induced by the localized mode, but caused by asymmetric mechanical stress, electric field, or some other external physical action, the abovementioned hypersensitivity will be equally strong in the setting (b) and (c). This effect can be used in switches, modulators and sensors.

Consider an asymmetric 1D photonic crystal (PC) consisting of two lossless quarter-wavelength Bragg gratings (BG) at the wavelength $\lambda_r = 4cm$. The constitutive components are different for each grating, as shown in Fig. 1. The refraction indices of the bilayers are $n_1 = 3.16$, $n_2 = 1$ and $n_3 = 1.5$, $n_4 = 4.74$ for the left BG (LBG) and the right BG (RBG), respectively. The periodic modulation of the index of refraction of each grating is engineered in such a way that both of them have the same band-gap structure, which is just a matter of convenience. At the interface between the two gratings an asymmetric cavity is created. The cavity consists of the two layers $n_1, n_4$ and a thin metallic cobult (Co) nanolayer placed between them. The Co layer has thickness $l_C = 0.18\mu m \ll \lambda_r$ and permittivity $\varepsilon_C(f) = i\frac{8.62}{f} \times 10^{16} Hz$ where $f$ is the frequency. Under typical circumstances the permittivity of each of the two layers of the cavity is affected differently by an external perturbation. For example the left layer $n_1$ can be more sensitive to high-level radiation than the right layer. The structure supports a localized mode with a frequency $f_r$ located in the middle of the photonic band-gap and the nodal point coinciding with the Co nanolayer.

The transmission $\mathcal{T}$, reflection $\mathcal{R}$, and absorption $\mathcal{A}$ are calculated via the transfer matrix approach. The latter connects the amplitudes of forward and backward propagating waves on the left and the right domains outside of the PC. At the $j-th$ layer inside the structure, and also outside of the PC, a time-harmonic field of angular frequency $\omega$ satisfies the Helmholtz equation:

$$\frac{d^2 E(z)}{dz^2} + \left(\frac{\omega}{c}\right)^2 \varepsilon E(z) = 0 \qquad (1)$$

where $\varepsilon = \varepsilon_j = n_j^2$ is the permittivity of the $j$-th layer ($\varepsilon = 1$ for the vacuum), and $c$ is the speed of light. At the $j-th$ layer, Eq. (1) admits solutions of the form $E^{(j)}(z) = E_f^{(j)} e^{in_j k z} + E_b^{(j)} e^{-in_j k z}$, where $k = \frac{\omega}{c}$ is the wavevector at the vacuum. Outside the PC, Eq. (1) admits the solution $E^{(L/R)}(z) = E_f^{(L/R)} e^{ikz} + E_b^{(L/R)} e^{-ikz}$. The continuity of the field and its derivative at the interface between two layers (or a layer and the vacuum) can be expressed in terms of the total transfer matrix $\mathcal{M}$ which connects the forward and backward amplitudes on the left (L) and right (R) of the PC:

$$\begin{pmatrix} E_f^{(R)} \\ E_b^{(R)} \end{pmatrix} = \begin{pmatrix} \mathcal{M}_{11} & \mathcal{M}_{12} \\ \mathcal{M}_{21} & \mathcal{M}_{22} \end{pmatrix} \begin{pmatrix} E_f^{(L)} \\ E_b^{(L)} \end{pmatrix}, \quad \mathcal{M} = \prod_{j=0}^{N} \mathcal{M}_j \qquad (2)$$

where $N$ is the total number of layers. The single-layer transfer matrix $\mathcal{M}_j$ connects the field amplitudes of the $j-th$ and the $(j+1)-th$ layers i.e. $\left(E_f^{(j+1)}, E_b^{(j+1)}\right)^T = \mathcal{M}_j \left(E_f^{(j)}, E_b^{(j)}\right)^T$. Thus the transfer matrix approach allows us also to construct the field $E^{(j)}(z)$ at each layer, provided that appropriate scattering boundary conditions are imposed. The latter, for a left incident wave, take the form $\left(E_f^{(R)}, E_b^{(R)}\right)^T = (1,0)^T$. It is easy to show that $\mathcal{T} = \left|\frac{1}{\mathcal{M}_{22}}\right|^2$, $\mathcal{R} = \left|\frac{\mathcal{M}_{21}}{\mathcal{M}_{22}}\right|^2$, and $\mathcal{A} = 1 - \mathcal{T} - \mathcal{R}$ [6,7].

We start our analysis with the investigation of the transmission spectra of each of the two mirrors. Their dispersion relation $\omega(q)$ is calculated using the transfer matrix of one bilayer $\mathcal{M}^{ab} = \mathcal{M}_b \mathcal{M}_a$ [7]

$$\mathcal{M}^{ab} = \begin{pmatrix} A & B \\ B^* & A^* \end{pmatrix}, \quad A = e^{in_b k l_b} \left[\cos(l_a n_a k) + \frac{i}{2}\left(\frac{n_a}{n_b} + \frac{n_b}{n_a}\right) \sin(l_a n_a k)\right], \quad B = \frac{i}{2} e^{in_b k l_b} \left(\frac{n_a}{n_b} - \frac{n_b}{n_a}\right) \sin(l_a n_a k), \quad (3).$$

where the indices a and b indicate the layers 1 and 2 (3 and 4) associated with the LBG (RBG).

Propagating waves in each grating correspond to frequencies $\omega = kc$ for which [7]

$$Trace(\mathcal{M}^{ab}) = 2\cos(l_a n_a k)\cos(l_b n_b k) - \left(\frac{n_a}{n_b} + \frac{n_b}{n_a}\right) \sin(l_a n_a k) \sin(l_b n_b k) \leq 2\cos(ql), \qquad (4)$$

where the total width of the bilayer $l = l_a + l_b$ defines the periodicity of the LBG (for $a = 1, b = 2$) or the RBG (for $a = 3, b = 4$). Direct inspection of Eq. (4) indicates that the dispersion relations $\omega_{LBG}(q)$ and $\omega_{RBG}(q)$ are identical as long as $\frac{n_1}{n_2} = \frac{n_3}{n_4}$ and $\frac{l_1}{l_2} = \frac{l_3}{l_4}$. In a finite photonic structure, both LBG and RBG will share the same band-gap structure of the transmission spectrum, as long as these conditions are satisfied. Below we consider that each BG consists of five (quarter-wavelength) bilayers.

In Fig. 1b we show the transmission spectrum $\mathcal{T}(f)$ of our PC for $\Delta\varepsilon = 0$. The position of the band-edges is described by Eq. (4). Moreover a resonant mode with $\mathcal{T}(f_r) \approx 1$ at resonance frequency $f_r \approx 7.5 GHz$, in the middle of the band-gap, has been created. The resonant mode is localized at the vicinity of the defect cavity and decays exponentially inside the two mirrors due to destructive interferences from the layers (blue profile at Fig. 1). The electric field $E(z)$ has a nodal point at the position of the metallic layer (blue profile at Fig. 1a). Thus the resonant localized mode is unaffected by the presence of the lossy layer and the entire PC is completely transparent at $f = f_r$ (see Fig. 1b). Furthermore, the lack of mirror symmetry ensures that the ASD occurs only for the resonance mode $f_r$. For all other (Fabry-Perot) resonances with frequencies $f \neq f_r$ the electric field distribution has finite amplitude at the position of the metallic layer leading to large reflection $\mathcal{R}(f) \approx 1$ (see discussion below) and vanishing transmission $\mathcal{T}(f) \approx 0$.

Moreover, any small perturbation (e.g., due to heating), which will change the permittivity of any of the two layers of the defect cavity (e.g., the left one) by $\Delta\varepsilon$, will engage immediately the metallic nanolayer and lift the ASD of the resonance localized mode. In other words, the electric field will no longer have a nodal point at the position of the metallic layer (see red profile at Fig. 1a). This will trigger various competing mechanisms. On the one hand it will increase the impendence mismatch and thus it will enhance the reflection. This mechanism is present whenever the electric field interacts with the metallic layer, even for $f \neq f_r$. At the same time it will lead to a non-monotonic behavior of absorption. One can understand qualitatively this mechanism by analyzing the transport from a lossy $\delta-$like layer with permittivity $\varepsilon(z) = i\gamma\delta(z)$. We have that:

$$\mathcal{T}_\delta(\omega) = \left(\frac{2}{2+k\gamma}\right)^2; \quad \mathcal{R} = \left(\frac{k\gamma}{2+k\gamma}\right)^2; \quad \mathcal{A} = \frac{4k\gamma}{(2+k\gamma)^2} \qquad (5)$$

which indicates that the layer is a source of absorption but at the same time a way to enhance reflection. Also note that the absorption is non-monotonic function of the tangent loss parameter $\gamma$. Rather it takes its maximum value $\mathcal{A}_{max} = 0.5$ at $\gamma = \frac{2}{k}$.

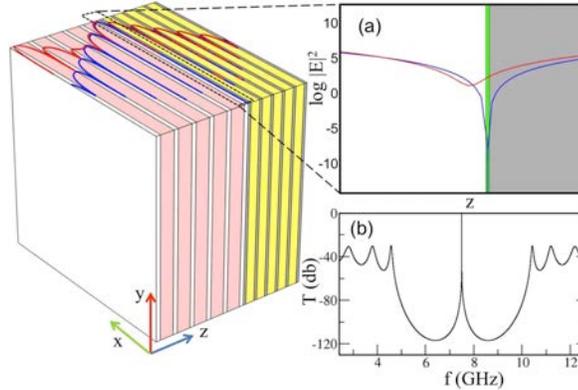

*Fig. 1. The proposed PC consisting of two BG with different constitutive bi-layers (white and orange for the LBG and grey and yellow for the RBG). The dashed box indicates the position of the asymmetric defect cavity at the interface of the two mirrors. (a) A magnification of the asymmetric cavity. It consists of (i) a white layer, (ii) a gray layer and (iii) a thin metallic nanolayer (green). We also report the electric field profile inside the PC at resonant frequency ($f \approx 7.50GHz$) for the unperturbed structure (blue), and the profile (red) at peak transmission $f \approx 7.46GHz$, for a 2% relative increase of the permittivity of the (white) layer of the defect cavity. The resonant mode is spoiled in the latter case. (b) The transmission vs. the frequency f for $\Delta\varepsilon = 0$. The resonance pick is associated with the blue electric field profile.*

The second mechanism applies only for resonant transport. In this case, the bulk losses due to the strong interaction of the electric field with the metallic layer compete with the losses due to the leakage from the boundaries of the structure. The former are proportional to the field intensity at the position of the lossy defect while the latter depend on the coupling of the resonant mode to the free space via the boundary of the PC. As $\Delta\varepsilon$ increases, the bulk losses overrun the losses due to the boundary leakage and eventually spoil the resonance (see red electric field in Fig. 1). Thus, photons do not dwell in the resonant mode and therefore cannot be absorbed by the metallic layer i.e. the absorption $\mathcal{A}(f_r)$ diminishes while $\mathcal{R}(f_r) \approx 1$, and $\mathcal{T}(f_r) \approx 0$.

In Fig. 2 we report the transmission $\mathcal{T}$, reflection $\mathcal{R}$, and absorption $\mathcal{A}$ of our PC, versus frequency and versus the relative permittivity change $\Delta\varepsilon/\varepsilon_1$ occurring at a single (left of the metallic nanolayer) layer. For small $\Delta\varepsilon/\varepsilon_1 \approx 0$, $\mathcal{T}(f_r) \approx 1$ (Fig. 2a) while $\mathcal{R} \approx 0, \mathcal{A} \approx 0$ respectively (Figs. 2b,c). As $\Delta\varepsilon$ increases, the absorption is initially increasing (see peak at $\Delta\varepsilon/\varepsilon_1 \sim 0.05\%$) but for $\Delta\varepsilon/\varepsilon_1 \gtrsim 1\%$ it starts decreasing reaching values as low as $-40\ dB$. For even larger $\Delta\varepsilon/\varepsilon_1$, the structure becomes completely reflective (see Fig. 2b).

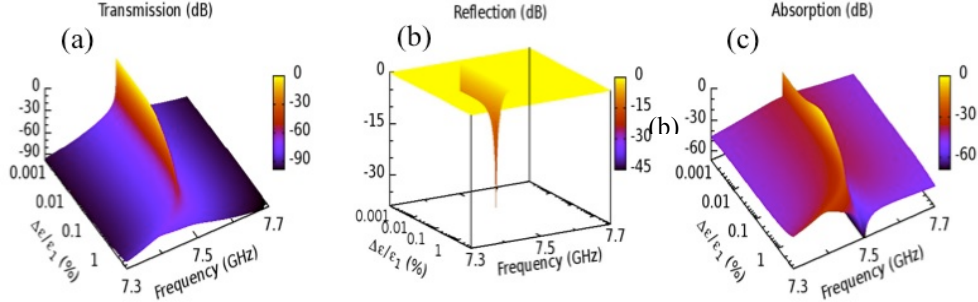

*Fig. 2. (a) Transmission (in dB's) versus frequency $f$ and percentage increase in permittivity change $\Delta\varepsilon/\varepsilon_1$. The reported frequency window corresponds to a part of the band-gap around the resonant frequency. For $\Delta\varepsilon/\varepsilon_1 \ll 1\%$, $\mathcal{T}(f_r) \approx 1$ at resonance frequency $f_r \approx 7.5$. For higher $\Delta\varepsilon/\varepsilon_1$, the transmission drops to values less than -40 dB. (b) The reflection $\mathcal{R}$ (in dBs) for the same frequency window and $\Delta\varepsilon/\varepsilon_1$. Low values of $\mathcal{R}$ ($\approx -10 dBs$), appearing only for small $\Delta\varepsilon/\varepsilon_1$ and at $f_r$, indicates small reflection. For $\Delta\varepsilon/\varepsilon_1 \approx 0.05\%$ the reflection becomes almost zero ($\approx -45 dB$) while the absorption gets its maximum value (see subfigure c). For larger $\Delta\varepsilon/\varepsilon_1 \geq 1\%$-values, our structure becomes reflective (i.e. $0 dBs$). (c) The absorption (in dB's). For $\Delta\varepsilon/\varepsilon_1 \approx 0$ the absorption $\mathcal{A}(f_r) \approx 0$ ($\approx -30 dBs$, see color coding). For small $\Delta\varepsilon/\varepsilon_1$ it increases and reaches a maximum at $\Delta\varepsilon/\varepsilon_1 \approx 0.05\%$. As $\Delta\varepsilon/\varepsilon_1$ increases further, $\mathcal{A}(f)$ decays abruptly and takes values smaller than -40 dBs.*

The hypersensitivity of the transport characteristics of our composite structure to small permittivity changes $\Delta\varepsilon$ finds various applications including sensors, switches, power modulators, etc. Here, however, we will discuss the advantages to implement our PC as an efficient energy limiter. These are devices that protect electromagnetic sensors from high-energy radiation, while at the same time they are transparent to low energy radiation [8,9]. Typically this protection is achieved via the absorption of the incident energy from the limiter, which turns opaque. At the same time this excessive energy overheats the limiter and leads to its self-destruction. Recently, however, the concept of reflective limiters has been introduced [10,11]. These structures are transmissive at low-energy incident pulses while they become highly reflective (and not absorbing) at high-energy pulses. Nevertheless, these proposals suffer from one drawback; the limiting action requires several orders of magnitude change of the permittivity of the lossy defect. Instead, our design requires changes of only a few percentage points in the permittivity of one composite layer in order to provide limiting action.

We consider the PC of Fig. 1. We further assume, for the sake of the discussion, that the left layer of the defect cavity [12] has a permittivity which depends on temperature ($T$) variations as $\varepsilon_1(T) = \varepsilon_1 + \Delta\varepsilon(T)$ where for simplicity we consider that $\Delta\varepsilon(T) = c_2 T$ [13]. The rate equation that determines the temporal behavior of the temperature $T(t)$ at the cavity is [11]:

$$\frac{d}{dt}T(t) = \frac{1}{C}\mathcal{A}(T)W_I(t) \qquad (6)$$

where $W_I(t) = |E_I(t)|^2$ is the incident pulse intensity, which is a given function of time, $C$ is the heat capacity, and $\mathcal{A}(T)$ is the temperature dependent absorption coefficient of the asymmetric cavity. A numerical integration of Eq. (6) (for a given pulse profile $W_I(t)$) allow us to evaluate the temperature $T(t)$ and further the permittivity variations $\Delta\varepsilon(T)$ which are reported in Fig. 3a as a percentage change $\Delta\varepsilon(T)/\varepsilon_1$. Then $\mathcal{T}(\omega)$, $\mathcal{R}(\omega)$ and $\mathcal{A}(\omega)$ are calculated using transfer matrices as a function of pulse duration $t$ (see Figs. 3b,c,d). We find that $\mathcal{A}(\omega)$ (Fig. 3b) initially increases and reaches some maximum value around $t \approx 0.1$ corresponding to very small permittivity changes $\Delta\varepsilon/\varepsilon_1 \lesssim 0.1\%$. Further increase of $\Delta\varepsilon(T)$ leads to an abrupt decay of $\mathcal{A}(\omega)$ for resonance frequencies to values smaller than -30 dB. The off-resonance values already have absorption that is below -60 dBs. At the same time the transmission (Fig. 3c) decays while the reflection (Fig. 3d) reaches unity. Therefore our photonic structure acts as a hypersensitive reflective microwave limiter- it will turn highly reflective within a broad frequency range for very small relative permittivity changes ~0.5%. This behavior has to be contrasted with the proposal of Ref. [11] where a limiting action is triggered only when the variation (due to heating) of the refraction index $\varepsilon_D = \varepsilon'_D + i\varepsilon''_D$ of a defect lossy layer, which is embedded in a Bragg grating, is many orders of magnitude. The outcome of these calculations are also reported in the inset of Fig. 3a by referring to $\mathcal{T}(f_r)$, $\mathcal{R}(f_r)$ and $\mathcal{A}(f_r)$ at resonance frequency versus the relative change of the permittivity [14]. We see that the reflective limiting action occurs when the permittivity changes of the lossy defect layer are more than seven orders of magnitude.

In conclusion, we have introduced PC designs with hypersensitive transport characteristics. These structures are based on asymmetric metal-dielectric defect cavities, which are embedded in a layered structure. When the developed nodal points of the associated defect localized mode coincide with the position of the metallic nano-layer of the defect cavity (accidental spatial degeneracy), the system demonstrates the phenomenon of induced transparency. However, even a small change in $\varepsilon'$ (and/or $\varepsilon''$) in one of the dielectric layers of the defect cavity will abruptly suppress the localized mode and render the entire structure highly reflective at all frequencies – not just at frequencies of the photonic band gap. Finally we have shown that these PC act as hypersensitive microwave or optical limiters with virtually unlimited protection frequency range.

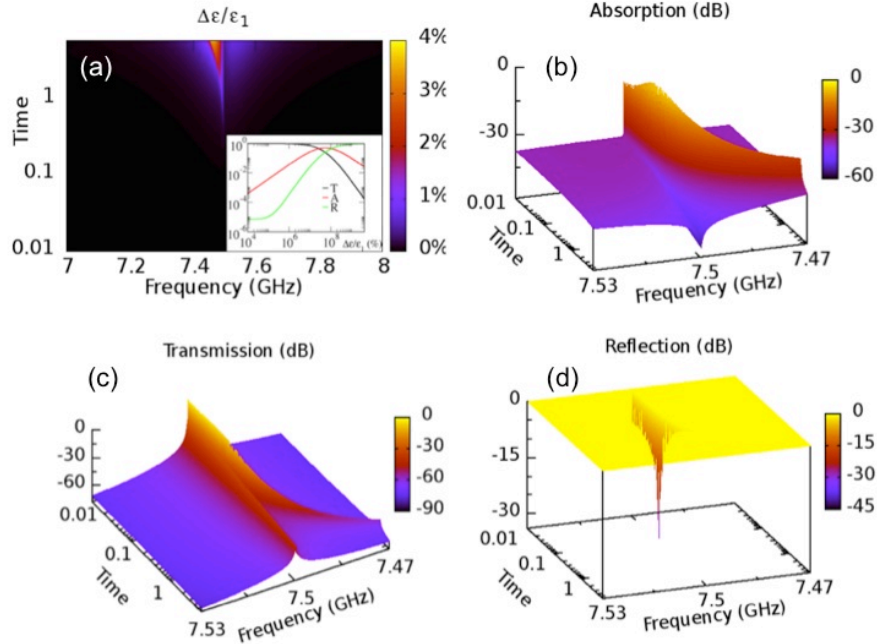

Figure 3: (a) A density plot of the percentage change of permittivity $\Delta\varepsilon/\varepsilon_1$ of the left layer of the defect cavity for a frequency range around the resonance localized mode, as a function of the pulse duration. Inset: the transmission T, reflection R and absorption A at resonance frequency versus $\Delta\varepsilon/\varepsilon_1$ for a reflective limiter of Ref. [11]. (b) The absorption; (c) the transmission; and (d) the reflection versus pulse duration, for a frequency window around the resonance mode.

*Acknowledgement:* We acknowledge AFOSR support from the portfolio of Dr. A. Nachman via LRIR09RY04COR Grant (I.V.) and MURI No. FA9550-14-1-0037 (E.M., T.K.). The effort of A.C. is supported from the AFOSR portfolio of Dr. A. Sayir.